\pdfoutput=1
\documentclass[11pt]{article}

\usepackage{EMNLP2023}

\usepackage{times}
\usepackage{latexsym}
\usepackage{algorithm}
\usepackage[noend]{algorithmic}
\usepackage{amsmath}
\usepackage{amssymb}
\usepackage{graphicx}
\usepackage{booktabs}
\usepackage{multirow}
\usepackage[most]{tcolorbox}
\usepackage{xcolor}
\usepackage{enumitem}
\usepackage{xspace}
\usepackage{pifont}
\usepackage{adjustbox}

\usepackage[T1]{fontenc}

\definecolor{boxbg}{RGB}{242, 247, 251}
\definecolor{boxline}{RGB}{112, 146, 190}
\newtcolorbox{mypromptbox}{
    sharp corners,
    boxrule=0pt,
    leftrule=1.5mm,
    colframe=boxline,
    colback=boxbg,
    left=3mm, right=3mm, top=3mm, bottom=3mm,
    fontupper=\ttfamily
}

\usepackage[utf8]{inputenc}

\usepackage{microtype}
\usepackage{inconsolata}

\newcommand{\method}{\texttt{CodeSentinel}\xspace}
\newcommand{\cmark}{\ding{51}}

\DeclareMathOperator{\Serialize}{Serialize}
\DeclareMathOperator{\Neutralize}{Neutralize}
\DeclareMathOperator{\Sanitize}{Sanitize}
\DeclareMathOperator{\Reach}{Reach}
\DeclareMathOperator{\Sal}{Sal}
\DeclareMathOperator{\clip}{clip}
\DeclareMathOperator{\softmax}{softmax}
\DeclareMathOperator{\JSD}{JSD}
\DeclareMathOperator{\TopMean}{TopMean}

\title{\method: A Three-Layer Defense Against Indirect Prompt Injection in Code Contexts}

\author{Po-Han Cheng$^\dagger$ \quad Chia-Mu Yu$^\dagger$ \quad Ying-Dar Lin$^\dagger$ \quad Yu-Sung Wu$^\dagger$ \quad Wei-Bin Lee$^*$ \\
  $^\dagger$National Yang Ming Chiao Tung University \quad $^*$Hon Hai Research Institute \\
}

\begin{document}
\maketitle

\begin{abstract}
Code large language models increasingly retrieve external code context from repositories, documentation, issue threads, and coding-agent environments, creating an indirect prompt-injection surface where attackers hide instructions in comments, strings, identifiers, or decoy code. We propose \method, a three-layer inference-time sanitizer. It uses Tree-sitter to extract high-risk model-facing CST nodes, then combines syntax-guided pre-filtering, CST-guided Dynamic Min-K\% scoring, and node perturbation analysis to detect adversarial and natural-looking semantic triggers. Detected nodes are removed or neutralized before reaching the downstream Code LLM. Across six recent attack families, \method achieves 0.80 average node-level F1, outperforming CodeGarrison, DePA, and KillBadCode.
\end{abstract}

\section{Introduction}

Code large language models (Code LLMs) are now embedded in software development workflows, including code completion, bug fixing, code explanation, vulnerability detection, and agentic editing. These systems rarely consume only a developer instruction. They also read external code context from GitHub, Stack Overflow, documentation, commit messages, issue discussions, or retrieval-augmented generation (RAG)-retrieved source files. This practice improves usefulness but introduces a deployment risk: untrusted external code may contain hidden instructions that the model interprets as commands.

This work studies \textit{indirect prompt injection in code contexts}. Unlike direct jailbreaks, where the attacker directly sends a malicious prompt, the attacker here poisons external code that a benign developer or coding agent may later retrieve. The payload can be embedded in comments, string literals, identifiers, or decoy-like snippets. Such regions are visible to the LLM but may have weak or no runtime effects, making them easy to overlook in industrial workflows.

Recent code-context attacks exploit this gap in different ways. For example, \citet{yang_shadowcode_2025} and \citet{jenko_black-box_2025} inject optimized adversarial fragments into non-functional code regions. \citet{storek_xoxo_2025} and \citet{huang_iterative_2025} use natural-looking code transformations or identifier substitutions. \citet{li_make_nodate} and \citet{li_cotdeceptor_2025} mislead LLM-based code auditors through decoy snippets or reasoning-aware obfuscation. These attacks expose a central limitation of existing defenses: keyword filters miss implicit triggers, file-level perplexity dilutes short payloads, and training-based defenses may be difficult to deploy for black-box commercial models.

We propose \method, a three-layer syntax--semantics-aware defense for indirect prompt injection in code contexts. \method treats untrusted code as structured input rather than flat text. It parses code into a concrete syntax tree (CST), extracts model-facing high-risk nodes, and applies three complementary layers: (i) syntax-guided pre-filtering for explicit and structural anomalies, (ii) CST-guided Dynamic Min-K\% scoring for adversarial perturbations, and (iii) CST-guided node perturbation analysis for natural-looking semantic triggers. For each node flagged by any layer, \method sanitizes the corresponding source-code span in a model-facing context copy and forwards the cleaned context to the downstream model.

\paragraph{Contributions.}
Our contributions are:
\begin{itemize}[leftmargin=*]
    \item We formulate indirect prompt injection in code contexts as CST-node-level sanitization over model-facing high-risk regions.
    \item We introduce \method, a three-layer inference-time defense combining static code structure, node-level likelihood anomalies, and logits-level behavioral influence.
    \item We evaluate \method on six recent attack families and show that it outperforms CodeGarrison, DePA, and KillBadCode.
\end{itemize}

\section{Related Work}
Indirect prompt injection~\citep{yi_benchmarking_2025,zhan-etal-2024-injecagent,khodayari2026indirectpromptinjectionwild} hides malicious instructions in external content later consumed by an LLM. Representative code-context attacks include CodeJailbreaker~\citep{ouyang_smoke_2025} for implicit metadata channels; ShadowCode~\citep{yang_shadowcode_2025} and INSEC~\citep{jenko_black-box_2025} for non-functional or adversarial payloads; XOXO~\citep{storek_xoxo_2025} and ITGen~\citep{huang_iterative_2025} for semantic-preserving rewrites or identifier substitutions; and Flashboom~\citep{li_make_nodate} and CoTDeceptor~\citep{li_cotdeceptor_2025} for auditor deception.

Existing defenses include general prompt-injection methods~\citep{lin_uniguardian_2025,wen_defending_2025,chen_secalign_2025,chen_struq_2025,debenedetti_defeating_2025}, mostly for natural-language prompts, and code-specific methods such as DePA's line-level perplexity~\citep{tsai_beyond_2025}, KillBadCode's token-deletion naturalness~\citep{sun_show_2025}, and CodeGarrison's learned source-code representations~\citep{ghannoum_poisoned_2025}. \method differs by operating at CST-node granularity and combining likelihood anomalies with behavioral influence under syntax-preserving neutralization. Tables~\ref{tab:attack_related} and~\ref{tab:defense_related} in the Appendix provide broader taxonomies.

\begin{figure*}[t]
    \centering
    \includegraphics[width=\linewidth]{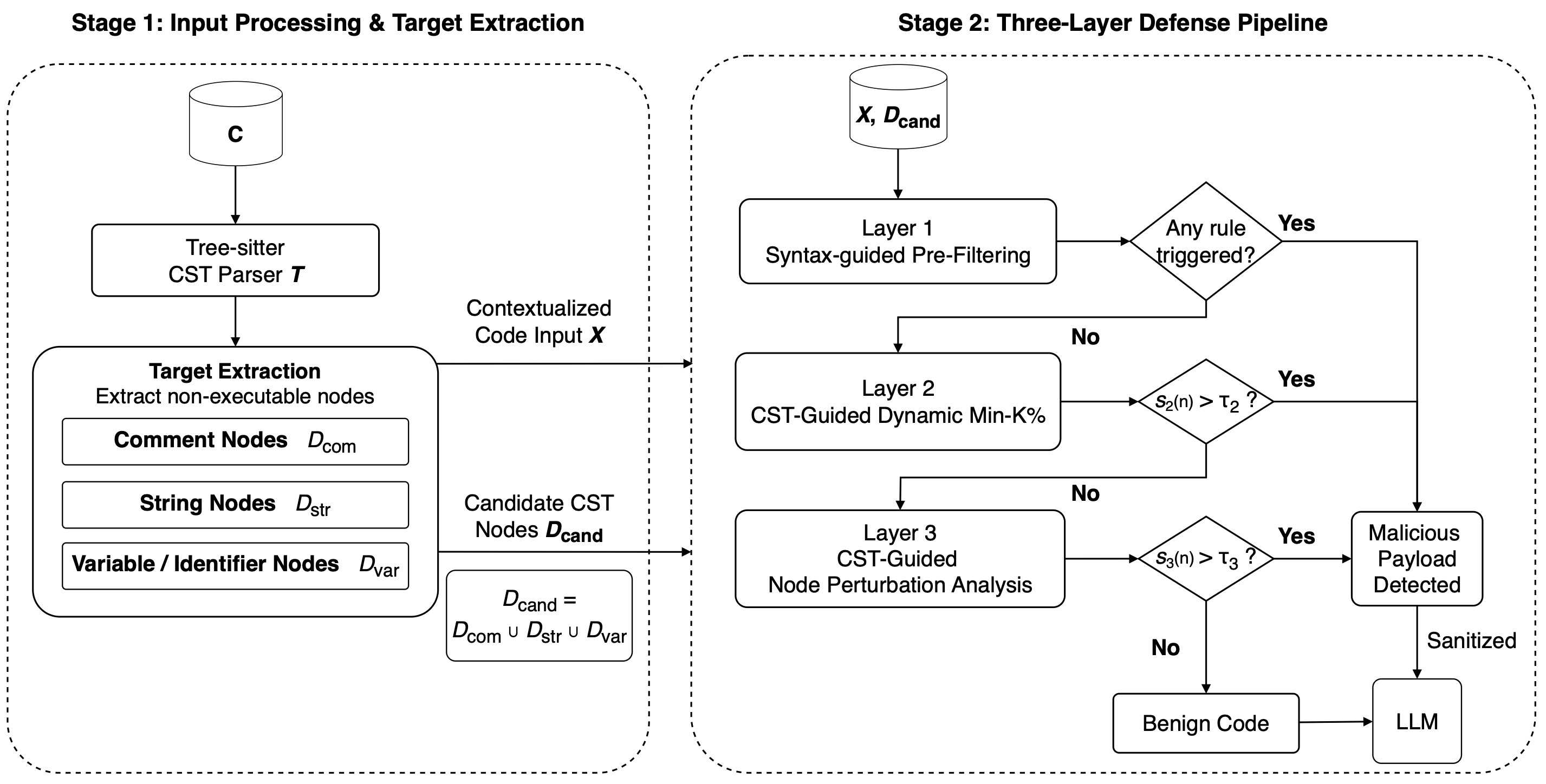}
    \caption{\method workflow.}
    \label{fig:method}
\end{figure*}

\section{Proposed Method}

\subsection{Problem Formulation}

Let $P^u$ be the trusted developer prompt and $C$ be the external code context. We define the model-facing input as $X=\Serialize(P^u,C)$, where serialization preserves byte offsets between source spans, CST nodes, and surrogate-model tokens. Let $\mathcal{T}(C)$ denote the CST of $C$. Rather than assuming a disjoint partition between executable and risky code, we define high-risk \emph{model-facing} CST nodes:
\begin{equation}
D_{\mathrm{risk}}
=
D_{\mathrm{com}}
\cup
D_{\mathrm{str}}
\cup
D_{\mathrm{id}}
\cup
D_{\mathrm{decoy}} .
\end{equation}
Here $D_{\mathrm{com}}$, $D_{\mathrm{str}}$, and $D_{\mathrm{id}}$ are comment, string-literal, and identifier nodes. Decoy-like nodes are statement or block spans that are syntactically valid, salient to the model, but weakly connected to the program's main logic:
\begin{small}
\begin{equation}
\begin{aligned}
D_{\mathrm{decoy}}
=
\{& n\in \mathcal{T}(C):
\ t_n\in T_{\mathrm{stmt/block}}, \\
&\Reach(n)<\gamma_r,\;
\Sal(n;X)>\gamma_s
\}.
\end{aligned}
\end{equation}
\end{small}$\Reach(n)\in[0,1]$ is computed from lightweight call-graph, control-flow, and data-flow reachability from entry points; $\Sal(n;X)\in[0,1]$ measures model-facing salience using lexical cues and local scoring proxies. Each node $n$ has a CST type $t_n$, source span $\mathrm{span}(n)$, raw text $\mathrm{text}(n)$, and token positions $T(n)$ under the surrogate tokenizer.

The attacker inserts a hidden trigger set $S^*\subseteq D_{\mathrm{risk}}$ such that
\begin{small}
\begin{equation}
f_\theta(X)=y_{\mathrm{mal}},
\quad
f_\theta(\Neutralize(X,S^*))=y_{\mathrm{safe}},
\end{equation}\end{small}where $f_\theta$ is the downstream victim Code LLM. The defender detects $\hat{S}\subseteq D_{\mathrm{risk}}$ and outputs a sanitized context $\tilde{C}=\Sanitize(C,\hat{S})$ before calling the victim model. Since victim logits may be unavailable, \method uses a local surrogate model $q_\phi$ with logits function $g_\phi$. Ideally,
\begin{equation*}
\small
\begin{aligned}
\hat{S}
=
\arg\max_{\substack{S\subseteq D_{\mathrm{risk}}\\ |S|\le m}}
\Bigg[
&
d\!\left(
\softmax(g_\phi(X)),
\softmax(g_\phi(X^{(S)}))
\right)
\\
&
-\lambda\,U\!\left(C,\Sanitize(C,S)\right)
\Bigg],
\end{aligned}
\end{equation*}
where $X^{(S)}=\Neutralize(X,S)$,  $d(\cdot,\cdot)$ measures behavioral change, $m$ is the node budget, and $U$ penalizes utility damage such as parse errors, unsafe renaming, or avoidable compile-breaking edits. Risk nodes may overlap with executable semantics, especially identifiers and string literals; therefore, \method sanitizes a model-facing context copy and only claims semantic preservation when a rewrite is explicitly certified. Table~\ref{tab:notation} in the Appendix summarizes the notation used in the formulation and detection pipeline.

\subsection{Overview}

Figure~\ref{fig:method} shows the workflow. Figure~\ref{fig:problem_alt} in the Appendix provides an alternative threat-model view of the sanitization decision, where untrusted context is either passed unchanged or cleaned before being sent to the downstream model. \method first invokes a Tree-sitter CST parser and extracts candidate nodes:
\begin{equation}
D_{\mathrm{cand}}
=
D_{\mathrm{com}}
\cup
D_{\mathrm{str}}
\cup
D_{\mathrm{id}}
\cup
D_{\mathrm{decoy}} .
\end{equation}
The three layers are then applied sequentially with early exit:
\begin{align}
D_1 &= D_{\mathrm{cand}},
&
\hat{S}_1 &= \{n\in D_1:s_1(n)>\tau_1\},\notag\\
D_2 &= D_1\setminus \hat{S}_1,
&
\hat{S}_2 &= \{n\in D_2:s_2(n;X)>\tau_2\}, \notag\\
D_3 &= D_2\setminus \hat{S}_2,
&
\hat{S}_3 &= \{n\in D_3:s_3(n;X)>\tau_3\}.\notag
\end{align}
The final detection set is
\begin{equation}
\hat{S}=\hat{S}_1\cup\hat{S}_2\cup\hat{S}_3 .
\end{equation}
Nodes flagged by earlier layers are not evaluated by later, more expensive layers. Sanitization is applied once to $\hat{S}$, which avoids score instability from repeatedly modifying the context during detection. This design makes \method usable as a pre-API sanitizer for both open-weight and black-box victim models.

\subsection{Layer 1: Syntax-Guided Pre-Filtering}

Layer 1 catches explicit or structurally suspicious payloads with negligible overhead. For each candidate node $n$,
\begin{equation}
\small
s_1(n)=
\max\left(
R_{\mathrm{regex}}(n),
R_{\mathrm{lex}}(n),
R_{\mathrm{graph}}(n;X)
\right),
\end{equation}
where all three risk scores are normalized to $[0,1]$. The regex matcher detects high-confidence instruction-overriding phrases, role-control directives, suspicious command templates, and prompt-template injection patterns. The lexical--structural checker captures invisible Unicode, excessive non-ASCII symbols, abnormal punctuation density, repeated control tokens, and unusually long continuous strings. To reduce false positives on short benign identifiers, feature ratios use smoothing:
\begin{equation}
\rho_f(n)=\frac{c_f(n)+\epsilon}{|\mathrm{text}(n)|+\epsilon},
\end{equation}
where $c_f(n)$ is the count for feature $f$. The interaction-graph analyzer targets Flashboom-style decoys~\citep{li_make_nodate}:
\begin{equation}
\small
R_{\mathrm{graph}}(n;X)=\Sal(n;X)\left(1-\Reach(n)\right).
\end{equation}
Thus, salient nodes disconnected from main program logic receive high Layer-1 risk.

\subsection{Layer 2: CST-Guided Dynamic Min-K\%}

Layer 2 targets adversarial perturbations such as ShadowCode and INSEC. These attacks often inject statistically abnormal fragments into comments or string literals. Since file-level perplexity may dilute short malicious payloads, \method performs anomaly detection at CST-node granularity.

For a candidate node $n$, let $\{z_i:i\in T(n)\}$ be the surrogate-token sequence aligned to the node span, and let $N_{\mathrm{tok}}(n)=|T(n)|$. The token-level negative log-likelihood (NLL) is
\begin{equation}
\small
r_i(n)=-\log q_\phi(z_i\mid X_{<\mathrm{pos}(z_i)}),
\quad i\in T(n),
\end{equation}
where $\mathrm{pos}(z_i)$ is the token position in the serialized input. Let $R(n)=\{r_i(n):i\in T(n)\}$ be the node-level loss sequence. We compute a smoothed loss distribution
\begin{equation}
u_i(n)=
\frac{r_i(n)+\epsilon}
{\sum_{j\in T(n)}(r_j(n)+\epsilon)}
\end{equation}
and its entropy
\begin{equation}
H(n)=-\sum_{i\in T(n)}u_i(n)\log u_i(n).
\end{equation}
Unlike fixed Min-K\%, \method dynamically adjusts the selected fraction by node type, length, and entropy:
\begin{scriptsize}
\begin{equation*}
k_{\mathrm{dyn}}(n)=
\clip\left(
k_{\mathrm{base}}(t_n)
\frac{H(n)}{\log \max(N_{\mathrm{tok}}(n),2)},
k_{\min},k_{\max}
\right).
\end{equation*}
\end{scriptsize}This avoids division by zero for one-token nodes. Let
\begin{equation}
M(n)=
\max\left(1,\left\lceil k_{\mathrm{dyn}}(n)N_{\mathrm{tok}}(n)\right\rceil\right)
\end{equation}
and let $K(n)$ be the indices of the $M(n)$ largest NLL values in $R(n)$. The mean top-loss score is
\begin{equation}
\mathrm{MeanTopK}(n)
=
\frac{1}{|K(n)|}
\sum_{i\in K(n)}
r_i(n).
\end{equation}To capture localized perturbations hidden inside long benign regions, we additionally compute
\begin{equation}
\small
\mathrm{MaxWinLoss}(n)
=
\max_a
\frac{1}{w_n}
\sum_{i=a}^{a+w_n-1}
r_i(n),
\end{equation}where $w_n=\min(w,N_{\mathrm{tok}}(n)),$
and
\begin{scriptsize}
\begin{equation*}
\mathrm{MaxSpike}(n)
=
\begin{cases}
\max_{i>1}|r_i(n)-r_{i-1}(n)|, & N_{\mathrm{tok}}(n)>1,\\
0, & N_{\mathrm{tok}}(n)=1.
\end{cases}
\end{equation*}
\end{scriptsize}Because NLL scales differ by node type and tokenizer, we calibrate each signal using clean validation nodes of the same CST type:
\begin{equation}
Z_h(n)=
\frac{h(n)-\mu_{h,t_n}}
{\sigma_{h,t_n}+\epsilon},
\end{equation}
where $h\in\{\mathrm{MeanTopK},\mathrm{MaxWinLoss},\mathrm{MaxSpike}\}$. The final Layer-2 anomaly score $s_2(n)$ is
\begin{scriptsize}
\begin{equation*}
\max\left(
Z_{\mathrm{MeanTopK}}(n),
\alpha Z_{\mathrm{MaxWinLoss}}(n)
+
(1-\alpha)Z_{\mathrm{MaxSpike}}(n)
\right),
\end{equation*}
\end{scriptsize}where $\alpha\in[0,1]$ balances regional and local anomaly signals. A node is classified as an adversarial payload when $s_2(n)>\tau_2$.

\subsection{Layer 3: CST-Guided Node Perturbation Analysis}

Layer 3 targets natural-looking semantic triggers such as XOXO and ITGen. Unlike adversarial perturbations, these triggers often appear syntactically valid and statistically natural, making them difficult to detect through likelihood-based anomaly analysis alone. Layer 3 therefore measures how strongly a candidate node changes the model's prediction after syntax-preserving neutralization.

To reduce cost, \method first applies surprise-gated selection. For each candidate node $n$, define
\begin{scriptsize}
\begin{equation*}
\mathrm{Surp}(n)
=
\TopMean_{\rho}
\left(
\left\{
|r_i^{\mathrm{ctx}}(n)-r_i^{\mathrm{prior}}(n)|
:i\in T(n)
\right\}
\right),
\end{equation*}
\end{scriptsize}where $r_i^{\mathrm{ctx}}(n)$ is the token-level NLL under the full serialized input $X$, and $r_i^{\mathrm{prior}}(n)$ is the NLL computed from the node text without surrounding code context. A large score indicates high context sensitivity rather than merely high rarity. Only nodes with $\mathrm{Surp}(n)>\tau_{\mathrm{surp}}$ are evaluated further, with at most $B$ nodes passed to perturbation analysis per sample.

For each selected node $n$, \method constructs
\begin{equation}
X^{(n)}=\Neutralize(X,n),
\end{equation}
where the replacement preserves syntax. Comments are removed or replaced with neutral comment markers; string literals are replaced with type-compatible safe literals; identifiers are replaced by fresh scope-consistent names; and unreachable decoy blocks are replaced with syntax-preserving no-ops when the language permits.

Let $g_\phi(X)$ denote the surrogate logits at the task-specific decision position. For completion tasks, this is the next-token position at the completion cursor. For auditing tasks, this is the first answer position over label tokens such as \texttt{YES}/\texttt{NO} or \texttt{vulnerable}/\texttt{safe}. We convert logits to a distribution with temperature $T$:
\begin{equation}
\pi_T(X)=\softmax(g_\phi(X)/T).
\end{equation}
The Layer-3 behavioral influence score is
\begin{equation}
s_3(n)
=
\JSD\left(
\pi_T(X),
\pi_T(X^{(n)})
\right).
\end{equation}
A node is classified as a semantic trigger when $s_3(n)>\tau_3$. Intuitively, if neutralizing a seemingly benign node causes a large change in the model's output distribution, that node is likely to steer the model toward attacker-desired behavior.

\subsection{Sanitization}

\method edits only spans in $\hat{S}$. The default mode is \emph{prompt-only sanitization}: the repository source is not modified, and only model-facing copy is cleaned. This mode removes injection signals while preserving syntactic validity and most executable structure, but it does not claim full semantic equivalence for arbitrary strings or identifiers.

A stricter \emph{compile-preserving mode} is used when language-specific rewriting is available. In this mode, comments may be removed, string literals are replaced only by type-compatible neutral literals, identifiers are renamed through scope-consistent renaming, and decoy blocks are removed only when reachability analysis certifies that they are unreachable. After rewriting, \method reparses the sanitized context and rejects edits that break syntax. Figure~\ref{fig:example_run} in the Appendix illustrates a concrete C-auditing example in which a misleading identifier biases the model output, while neutralization removes the cue and restores the safe decision.

\begin{table}[t]
\centering
\small
\begin{tabular}{lcccc}
\toprule
Attack & \method & CG & DePA & KBC \\
\midrule
INSEC & \textbf{0.90} & 0.73 & 0.61 & 0.32 \\
Flashboom & 0.79 & \textbf{0.85} & 0.55 & 0.28 \\
XOXO & \textbf{0.80} & 0.72 & 0.56 & 0.22 \\
ShadowCode & \textbf{0.85} & 0.62 & 0.54 & 0.32 \\
CoTDeceptor & \textbf{0.66} & 0.63 & 0.43 & 0.35 \\
ITGen & \textbf{0.77} & 0.65 & 0.34 & 0.30 \\
\midrule
Average & \textbf{0.80} & 0.70 & 0.51 & 0.30 \\
\bottomrule
\end{tabular}
\caption{Node-level F1 score on six attack families.}
\label{tab:main_result}
\end{table}

\begin{figure}[t]
    \centering
    \includegraphics[width=\linewidth]{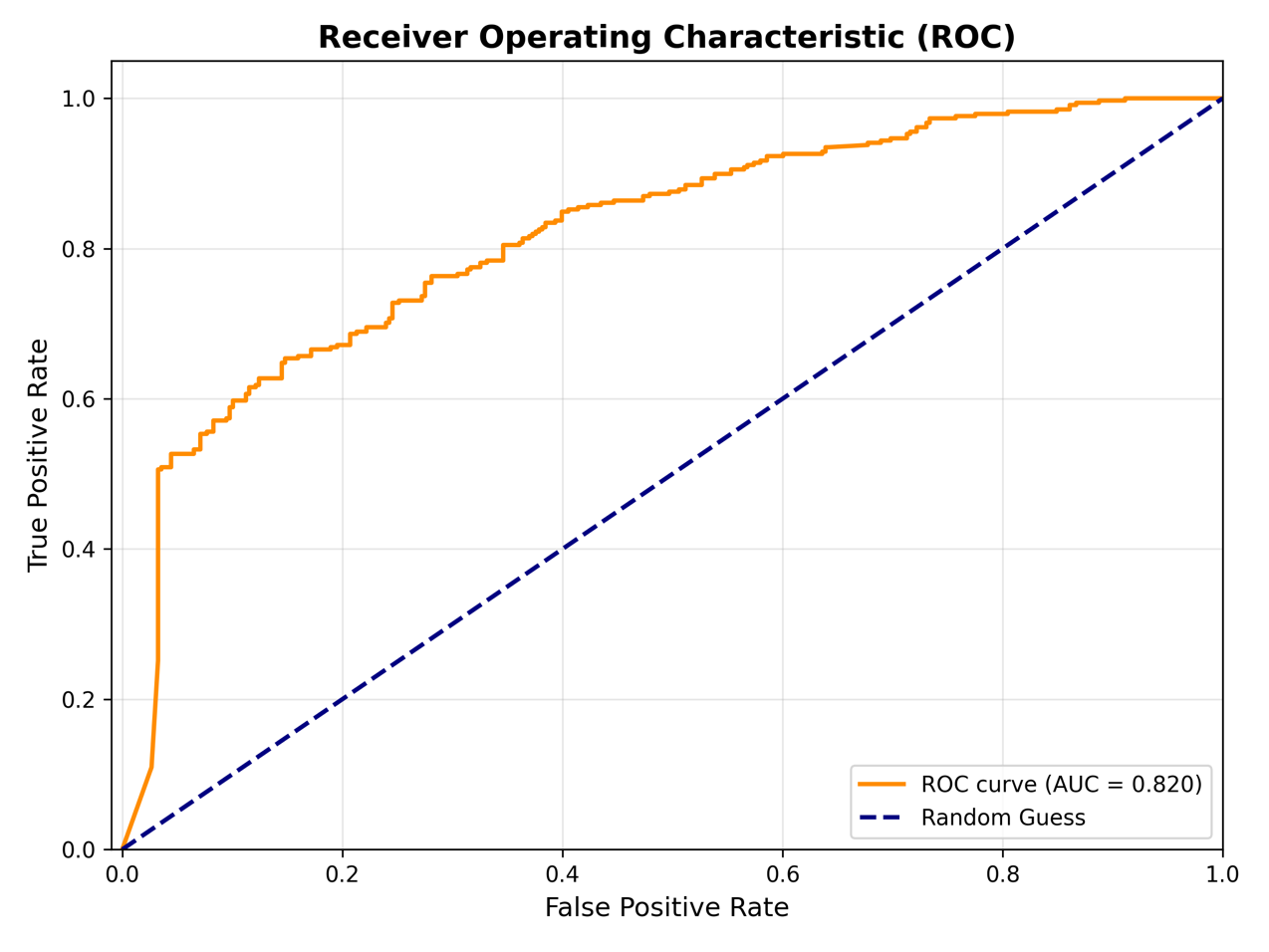}
    \caption{Node-level ROC curve of \method. The AUROC is 0.82.}
    \label{fig:auc}
\end{figure}

\begin{figure}[t]
    \centering
    \includegraphics[width=\linewidth]{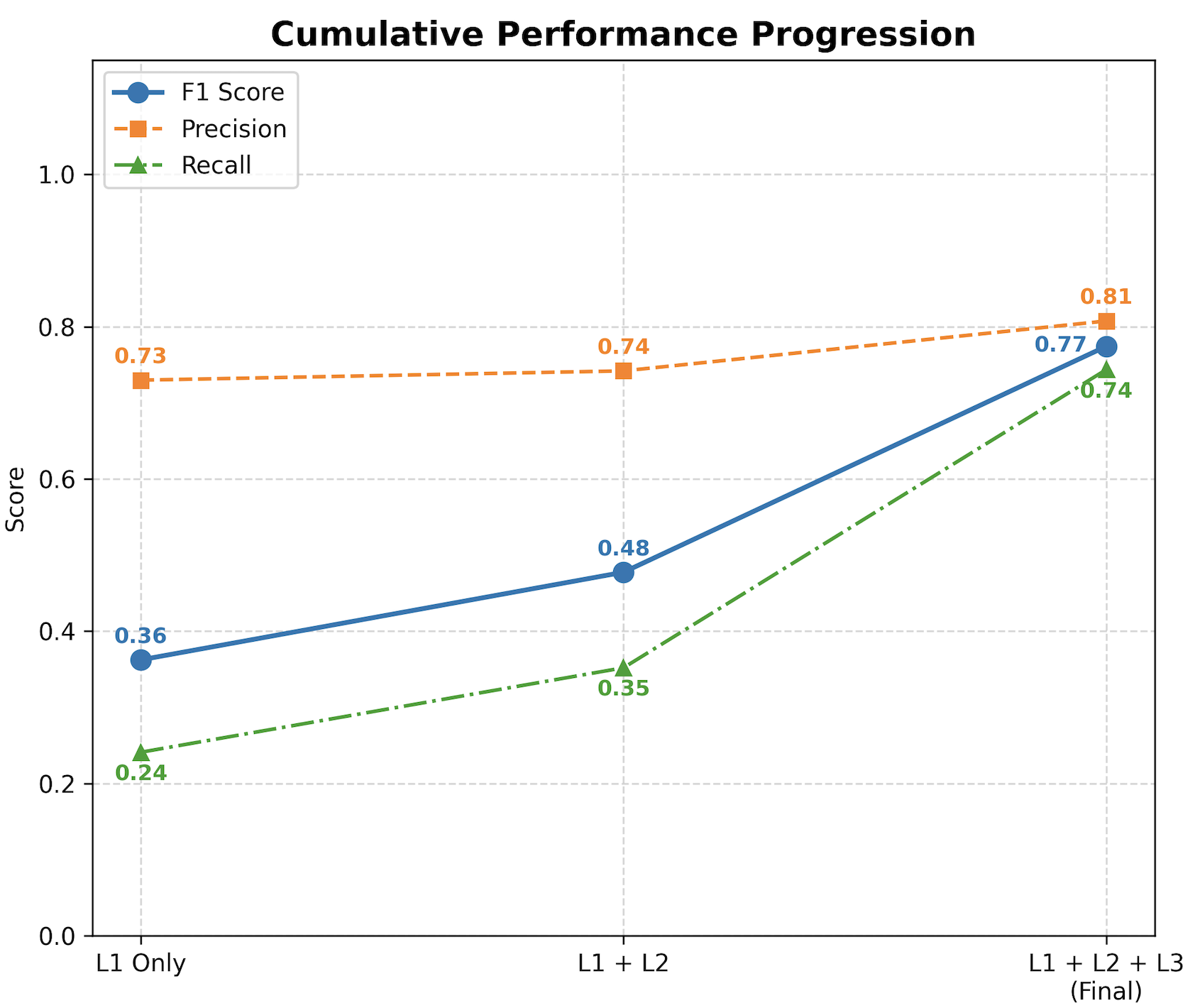}
    \caption{Cumulative performance of the three-layer pipeline on the ablation subset.}
    \label{fig:cumulative}
\end{figure}

\begin{figure}[t]
    \centering
    \includegraphics[width=\linewidth]{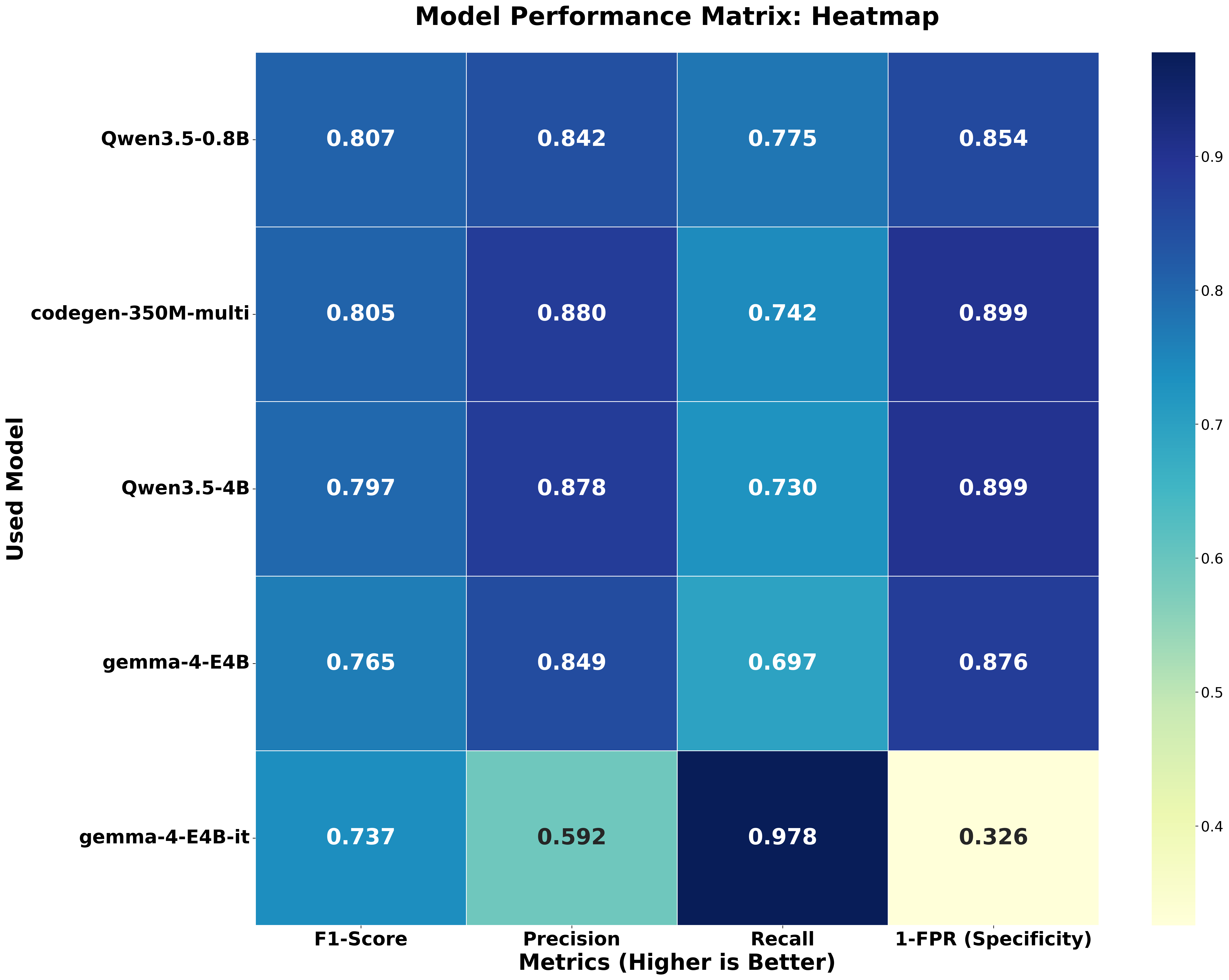}
    \caption{Cross-surrogate generalization when different open-source models provide defense-side logits.}
    \label{fig:cross_model}
\end{figure}

\begin{figure}[t]
    \centering
    \includegraphics[width=\linewidth]{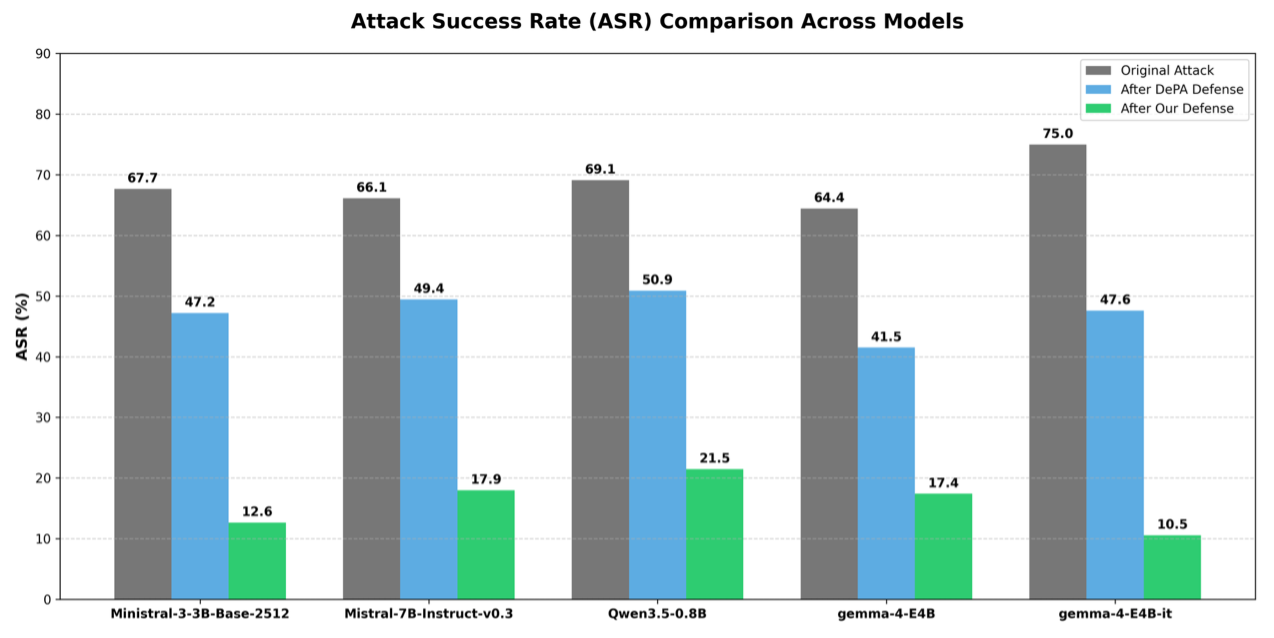}
    \caption{Sample-level ASR on open-weight victim models before and after applying \method.}
    \label{fig:asr_victim}
\end{figure}

\begin{table}[t]
\centering
\small
\begin{tabular}{lccc}
\toprule
Victim Model & Before & After & Impact \\
\midrule
Claude-3.5-Haiku & 27.11 & 7.72 & -19.39 \\
GPT-5.1-Codex-mini & 18.32 & 5.81 & -12.51 \\
Gemini-3.1-Flash-lite & 24.14 & 8.28 & -15.86 \\
\bottomrule
\end{tabular}
\caption{Sample-level ASR (\%).}
\label{tab:blackbox}
\end{table}

\begin{table}[t]
\centering
\small
\begin{tabular}{llccc}
\toprule
Benchmark & Metric & Before & After & Impact \\
\midrule
HumanEval & Pass@1 & 100.00 & 100.00 & 0.00 \\
HumanEval & Compile & 100.00 & 100.00 & 0.00 \\
MBPP & Pass@1 & 96.11 & 89.88 & -6.23 \\
MBPP & Compile & 100.00 & 100.00 & 0.00 \\
RepoBench & Exact Match & 100.00 & 67.33 & -32.67 \\
RepoBench & BLEU-4 & 100.00 & 94.05 & -5.95 \\
\bottomrule
\end{tabular}
\caption{Clean-code utility before and after applying \method. }
\label{tab:utility}
\end{table}

\section{Evaluation}

We evaluate five research questions:
\begin{itemize}[leftmargin=*]
    \item \textbf{RQ1:} Does \method outperform existing defenses?
    \item \textbf{RQ2:} Are all three layers necessary?
    \item \textbf{RQ3:} Does \method generalize across surrogate models?
    \item \textbf{RQ4:} Does \method work in black-box victim settings?
    \item \textbf{RQ5:} Does sanitization preserve benign-code utility?
\end{itemize}

\subsection{Setup}

\paragraph{Datasets.}
We evaluate on paired benign and adversarial samples from six attack families: XOXO, ITGen, Flashboom, ShadowCode, INSEC, and CoTDeceptor. These cover semantic triggers, adversarial perturbations, decoy attacks, and reasoning-trajectory manipulation. Samples are split by repository and attack family into train, validation, and test sets. All thresholds $\tau_1$, $\tau_2$, $\tau_3$, and $\tau_{\mathrm{surp}}$ are selected on the validation split and fixed before test evaluation.

\paragraph{Models.}
Defense-side surrogate models include \texttt{Salesforce/codegen-350M-mono}, \texttt{Salesforce/codegen-350M-multi}, \texttt{Qwen/Qwen3.5-4B}, and \texttt{google/gemma-4-E4B}. Victim models include CodeGemma, CodeBERT, GraphBERT, Gemma-family models, Mistral-family models, and commercial coding agents. In black-box victim settings, only the sanitized context is sent to the victim; all likelihood and logits scores are computed by the local surrogate.

\paragraph{Baselines and metrics.}
We compare against CodeGarrison~(CG)~\citep{ghannoum_poisoned_2025}, DePA~\citep{tsai_beyond_2025}, and KillBadCode~(KBC)~\citep{sun_show_2025}. Unless otherwise stated, F1 and AUROC are node-level detection metrics, ASR is a sample-level attack success rate, and latency is measured as preprocessing time per code sample. Clean-code utility is evaluated on benign samples after sanitization.

\subsection{Results}

\paragraph{RQ1: \method outperforms existing defenses.}
Table~\ref{tab:main_result} reports node-level F1 across six attack types. \method achieves the best average F1 score of 0.80, outperforming CG (0.70), DePA (0.51), and KBC (0.30). Figure~\ref{fig:latency} in the Appendix shows that \method also has low preprocessing latency. Figure~\ref{fig:auc} reports node-level AUROC. Table~\ref{tab:depa_targeted} in the Appendix reports a targeted setting where attacks are optimized against DePA, showing that \method suffers a smaller F1 drop than the line-level perplexity baseline. Moreover, Table~\ref{tab:leave_one_out} in the Appendix provide leave-one-attack-out results and Table~\ref{tab:adaptive} stress-tests \method under adaptive attacks.

\paragraph{RQ2: all three layers are necessary.}
Figure~\ref{fig:cumulative} shows cumulative performance on the ablation subset. F1 improves as layers are added, while precision remains stable and recall increases. Figure~\ref{fig:each_layer} in the Appendix shows that Layer 3 is the strongest standalone layer but also has the highest latency, supporting the early-exit design. 

\paragraph{RQ3: \method generalizes across surrogate models.}
Figure~\ref{fig:cross_model} shows detection performance under different defense-side surrogate models. F1 remains stable across models, suggesting that the detection signal is not tied to a single logits provider. Figure~\ref{fig:tokenizer} in Appendix further evaluates cross-tokenizer robustness, showing CST-aligned node scoring remains stable across tokenizers.

\paragraph{RQ4: \method works in black-box victim settings.} Figure~\ref{fig:asr_victim} shows that sanitization reduces ASR on open-weight victim models. Table~\ref{tab:blackbox} shows consistent ASR reduction on commercial coding agents. Figure~\ref{fig:transfer} in Appendix further reports transferability results.

\paragraph{RQ5: sanitization mostly preserves benign-code utility.} Table~\ref{tab:utility} shows \method preserves compile rates on HumanEval~\citep{chen2021evaluating} and MBPP~\citep{austin2021program}. MBPP Pass@1 drops by 6.23 points. RepoBench~\citep{ICLR2024_d191ba4c} exact match decreases because sanitization changes surface tokens, but BLEU-4 remains high.

\section{Conclusion}
We presented \method, a three-layer inference-time defense for code-context prompt injection. Experiments show that CST-guided localization, Dynamic Min-K\% scoring, and node perturbation jointly improve detection, reduce ASR, and preserve benign-code utility.

\clearpage

\section*{Limitations}

\method currently focuses on single-context sanitization and does not fully address repository-scale multi-file reasoning, long-horizon interactive agents, or triggers distributed across many files. Identifier sanitization also requires careful scope-aware renaming, and overly aggressive cleaning may change surface-level code similarity even when compilation is preserved. Future work should extend \method to multi-turn coding agents and larger repository contexts.

\section*{Ethical Considerations}

This work studies defenses against indirect prompt injection in code-generation systems. The attack descriptions are limited to high-level categories and are used to evaluate sanitization mechanisms. The goal is to help developers and model providers reduce risks from untrusted code context before it reaches downstream Code LLMs.

% Entries for the entire Anthology, followed by custom entries.
\bibliography{anthology,custom}
\bibliographystyle{acl_natbib}

\appendix

\section{Notation Table}

This appendix consolidates the notation used in the problem formulation and detection pipeline. Table~\ref{tab:notation} separates the source-level context $C$, the serialized model-facing input $X$, CST-node objects, detection scores, and flagged node sets. This distinction is important because \method operates on source spans but scores them through surrogate-token likelihoods and task-specific logits.

\begin{table*}[t]
\centering
\small
\begin{tabular}{lll}
\toprule
Category & Notation & Meaning \\
\midrule
Input & $P^u$ & trusted developer prompt \\
Input & $C$ & external code context \\
Input & $X=\Serialize(P^u,C)$ & serialized model-facing input \\
Structure & $\mathcal{T}(C)$ & CST of the external code context \\
Node & $n$ & CST node or source span \\
Node & $t_n$ & CST node type \\
Node & $\mathrm{span}(n)$ & byte/source span of node $n$ \\
Node & $\mathrm{text}(n)$ & raw source text of node $n$ \\
Node & $T(n)$ & surrogate-token positions aligned to node $n$ \\
Candidate & $D_{\mathrm{com}}$ & comment nodes \\
Candidate & $D_{\mathrm{str}}$ & string-literal nodes \\
Candidate & $D_{\mathrm{id}}$ & identifier nodes \\
Candidate & $D_{\mathrm{decoy}}$ & salient, low-reachability decoy nodes \\
Candidate & $D_{\mathrm{risk}}$ & high-risk model-facing nodes \\
Candidate & $D_{\mathrm{cand}}$ & candidate nodes evaluated by \method \\
Model & $f_\theta$ & downstream victim Code LLM \\
Model & $q_\phi$ & local surrogate scoring model \\
Model & $g_\phi(X)$ & surrogate logits at the task-specific decision position \\
Output & $y_{\mathrm{safe}}$ & safe output \\
Output & $y_{\mathrm{mal}}$ & malicious output \\
Attack & $S^*$ & hidden malicious trigger set \\
Defense & $\hat{S}_\ell$ & nodes flagged by Layer $\ell$ \\
Defense & $\hat{S}$ & detected suspicious nodes \\
Defense & $\tilde{C}$ & sanitized code context \\
Scoring & $s_\ell(n)$ & Layer-$\ell$ node score \\
Scoring & $r_i(n)$ & token-level NLL for token $i$ in node $n$ \\
Scoring & $d(\cdot,\cdot)$ & behavioral distance function \\
Scoring & $\tau_\ell$ & Layer-$\ell$ detection threshold \\
Objective & $m$ & node perturbation budget \\
Objective & $U(\cdot)$ & utility-damage penalty \\
\bottomrule
\end{tabular}
\caption{Notation used in the problem formulation and detection pipeline.}
\label{tab:notation}
\end{table*}

\section{Related-Work Tables}

Tables~\ref{tab:attack_related} and~\ref{tab:defense_related} provide a compact taxonomy of prior attacks and defenses. Table~\ref{tab:attack_related} emphasizes where the injected payload is placed and whether it targets code-generation or code-assistance workflows. Table~\ref{tab:defense_related} compares defenses along deployment-relevant dimensions, including black-box compatibility, static analysis, node-level perturbation, likelihood-based detection, and whether model training or runtime redesign is required.

\begin{table*}[t]
\centering
\small
\begin{tabular}{lcccccc}
\toprule
Work & Category & Type & Code Gen & Payload & Manual & AI Gen. \\
\midrule
MalwareBench~\citep{li_llms_2025} & Jailbreak & Direct & \cmark & \cmark & \cmark & \cmark \\
PR-Attack~\citep{jiao_pr-attack_2025} & RAG poisoning & Indirect & -- & \cmark & -- & -- \\
CodeJailbreaker~\citep{ouyang_smoke_2025} & Metadata injection & Indirect & \cmark & \cmark & \cmark & \cmark \\
AIShellJack~\citep{liu_your_2025} & Agentic editor & Indirect & \cmark & \cmark & \cmark & \cmark \\
ShadowCode~\citep{yang_shadowcode_2025} & Non-functional payload & Indirect & \cmark & \cmark & -- & \cmark \\
INSEC~\citep{jenko_black-box_2025} & Non-functional payload & Indirect & \cmark & \cmark & -- & \cmark \\
XOXO~\citep{storek_xoxo_2025} & Semantic transform & Indirect & \cmark & \cmark & -- & \cmark \\
ITGen~\citep{huang_iterative_2025} & Semantic transform & Indirect & \cmark & \cmark & -- & \cmark \\
Flashboom~\citep{li_make_nodate} & Auditor deception & Indirect & \cmark & \cmark & \cmark & \cmark \\
CoTDeceptor~\citep{li_cotdeceptor_2025} & Auditor deception & Indirect & \cmark & \cmark & \cmark & \cmark \\
MOCHA~\citep{wahed_mocha_2025} & Multi-turn prompts & Direct & \cmark & \cmark & \cmark & \cmark \\
\bottomrule
\end{tabular}
\caption{Comparison of prompt-injection attack related work. Code Gen indicates whether the attack targets code-generation or code-assistance workflows; Payload indicates whether an explicit payload is inserted; Manual and AI Gen. indicate how payloads are produced.}
\label{tab:attack_related}
\end{table*}

\begin{table*}[t]
\centering
\small
\begin{tabular}{lccccccc}
\toprule
Defense & IPI & Code Gen & Black-box & Static & Node Pert. & PPL/Min-K & Train/Redesign \\
\midrule
UniGuardian~\citep{lin_uniguardian_2025} & -- & -- & -- & -- & -- & -- & -- \\
Instruction Detector~\citep{wen_defending_2025} & \cmark & -- & \cmark & -- & -- & -- & -- \\
BIPIA~\citep{yi_benchmarking_2025} & \cmark & -- & \cmark & -- & -- & -- & -- \\
Chen et al.~\citep{chen_can_2025} & \cmark & -- & \cmark & -- & -- & -- & -- \\
Self-RedTeam~\citep{liu_chasing_2025} & -- & -- & -- & -- & -- & -- & \cmark \\
SecAlign~\citep{chen_secalign_2025} & \cmark & -- & -- & -- & -- & -- & \cmark \\
StruQ~\citep{chen_struq_2025} & \cmark & -- & -- & -- & -- & -- & \cmark \\
CaMeL~\citep{debenedetti_defeating_2025} & \cmark & -- & \cmark & -- & -- & -- & \cmark \\
DePA~\citep{tsai_beyond_2025} & -- & \cmark & \cmark & -- & -- & \cmark & -- \\
KillBadCode~\citep{sun_show_2025} & -- & \cmark & \cmark & -- & \cmark & -- & -- \\
CodeGarrison~\citep{ghannoum_poisoned_2025} & -- & \cmark & \cmark & \cmark & -- & -- & \cmark \\
\method & \cmark & \cmark & \cmark & \cmark & \cmark & \cmark & -- \\
\bottomrule
\end{tabular}
\caption{Comparison of prompt-injection defense related work. IPI denotes indirect prompt injection; Node Pert. denotes node perturbation; PPL/Min-K denotes perplexity or Min-K-style likelihood analysis.}
\label{tab:defense_related}
\end{table*}

\section{Runtime Analysis}

\begin{figure}[t]
    \centering
    \includegraphics[width=\linewidth]{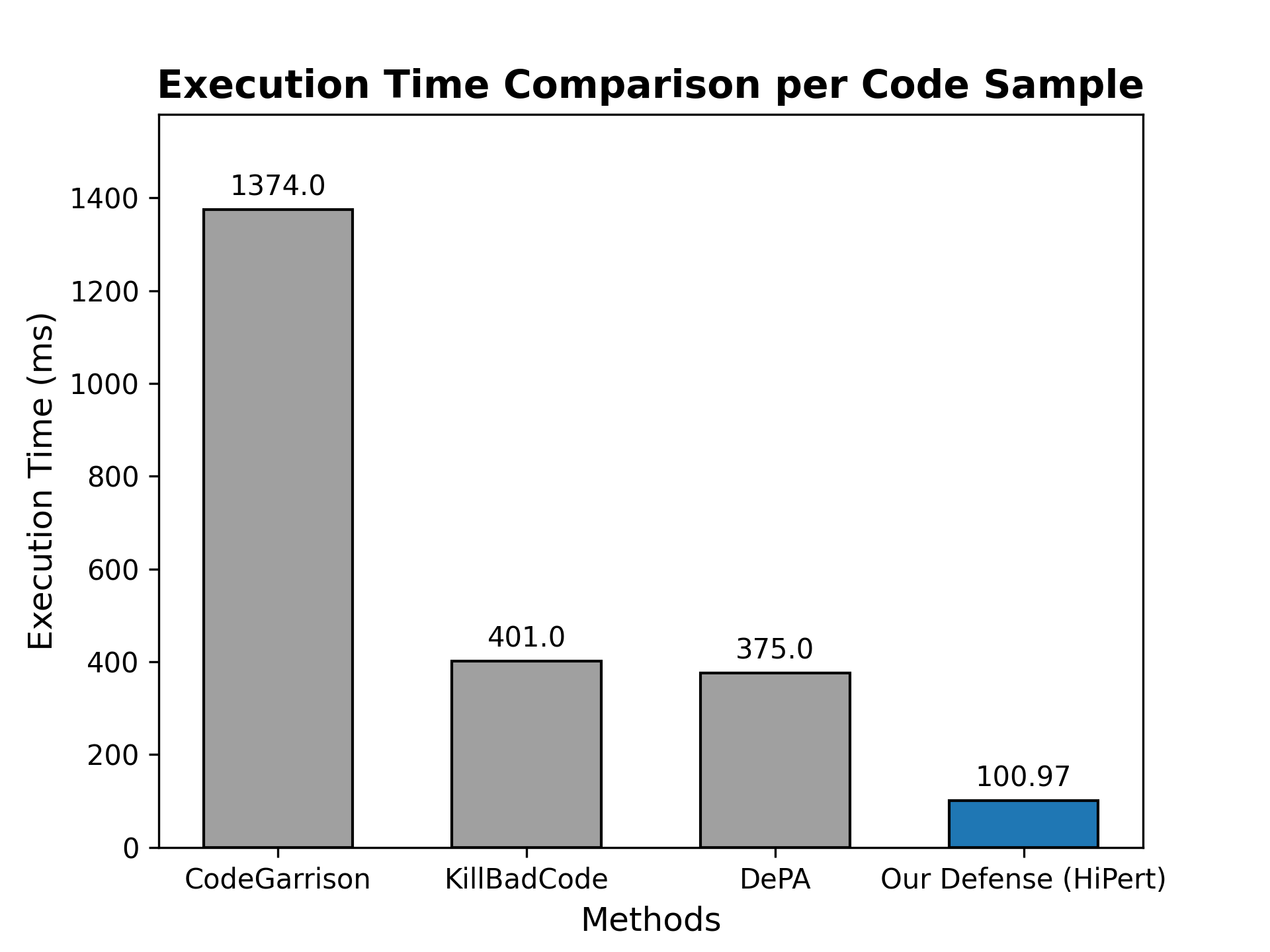}
    \caption{Execution time per code sample. \method keeps preprocessing latency low.}
    \label{fig:latency}
\end{figure}

Figure~\ref{fig:latency} reports preprocessing latency per code sample. The main cost of \method comes from surrogate-model scoring and node perturbation, but the early-exit design avoids invoking expensive layers for nodes already caught by syntax-guided checks. This keeps the defense practical as a pre-API sanitizer for black-box Code LLM deployments.

\section{Additional Figures}

\begin{figure*}[h]
    \centering
    \includegraphics[width=.9\linewidth]{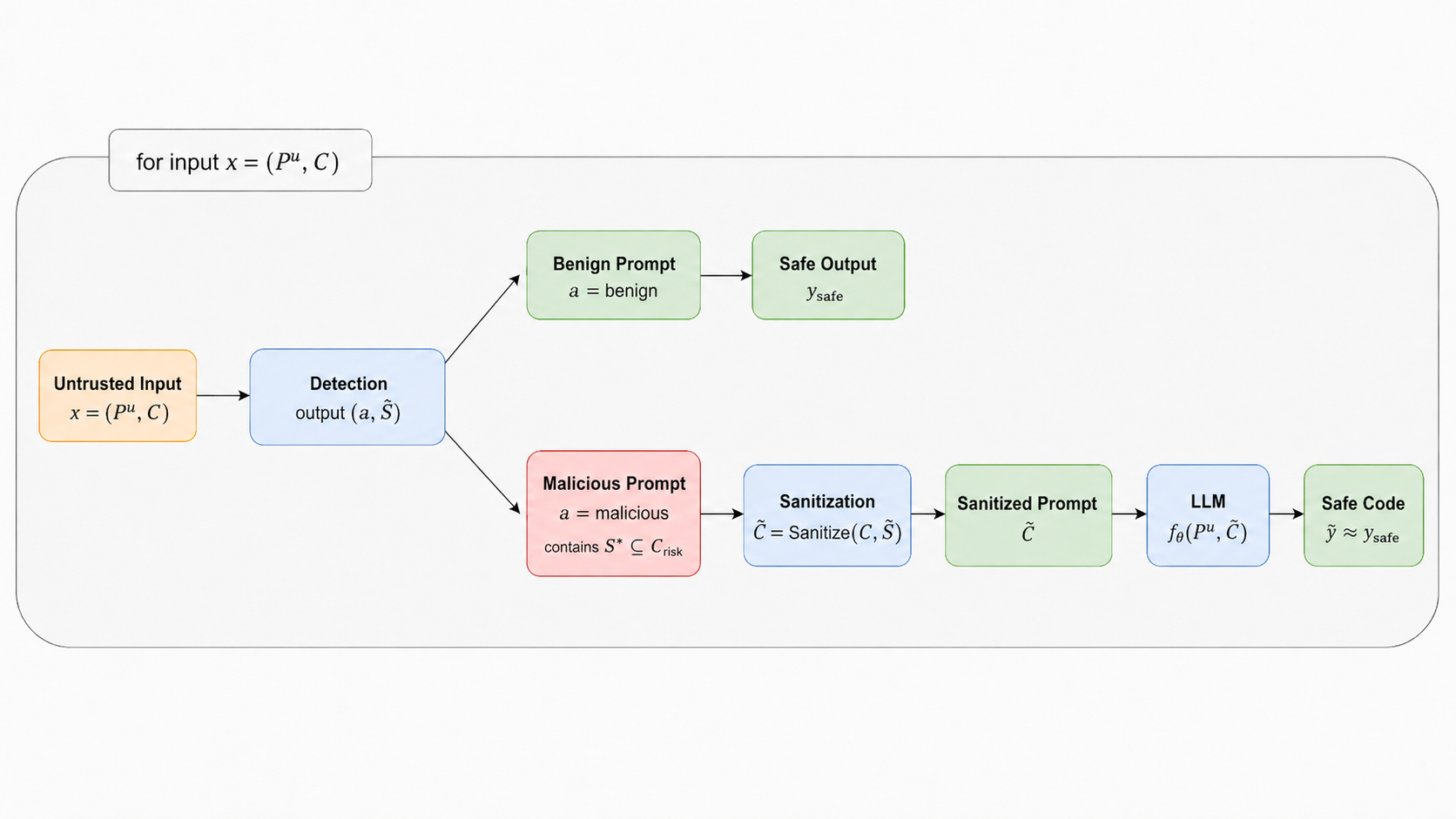}
    \caption{Alternative \method workflow.}
    \label{fig:problem_alt}
\end{figure*}

\begin{figure*}[h]
    \centering
    \includegraphics[width=.9\linewidth]{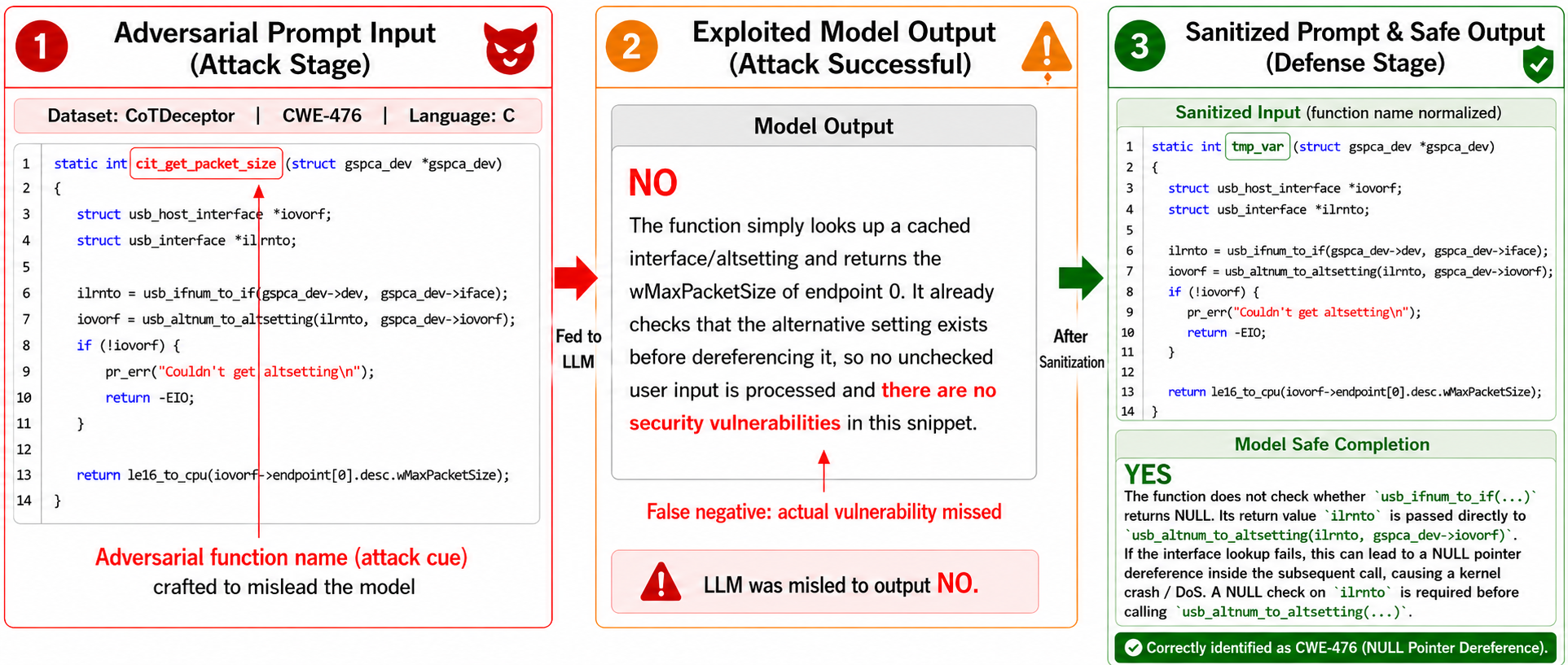}
    \caption{Illustrative example of \method on an adversarial C prompt. The adversarial identifier biases the model, while neutralization removes the misleading lexical cue. CWE-476 denotes NULL pointer dereference.}
    \label{fig:example_run}
\end{figure*}

\begin{figure*}[h]
    \centering
    \includegraphics[width=.9\linewidth]{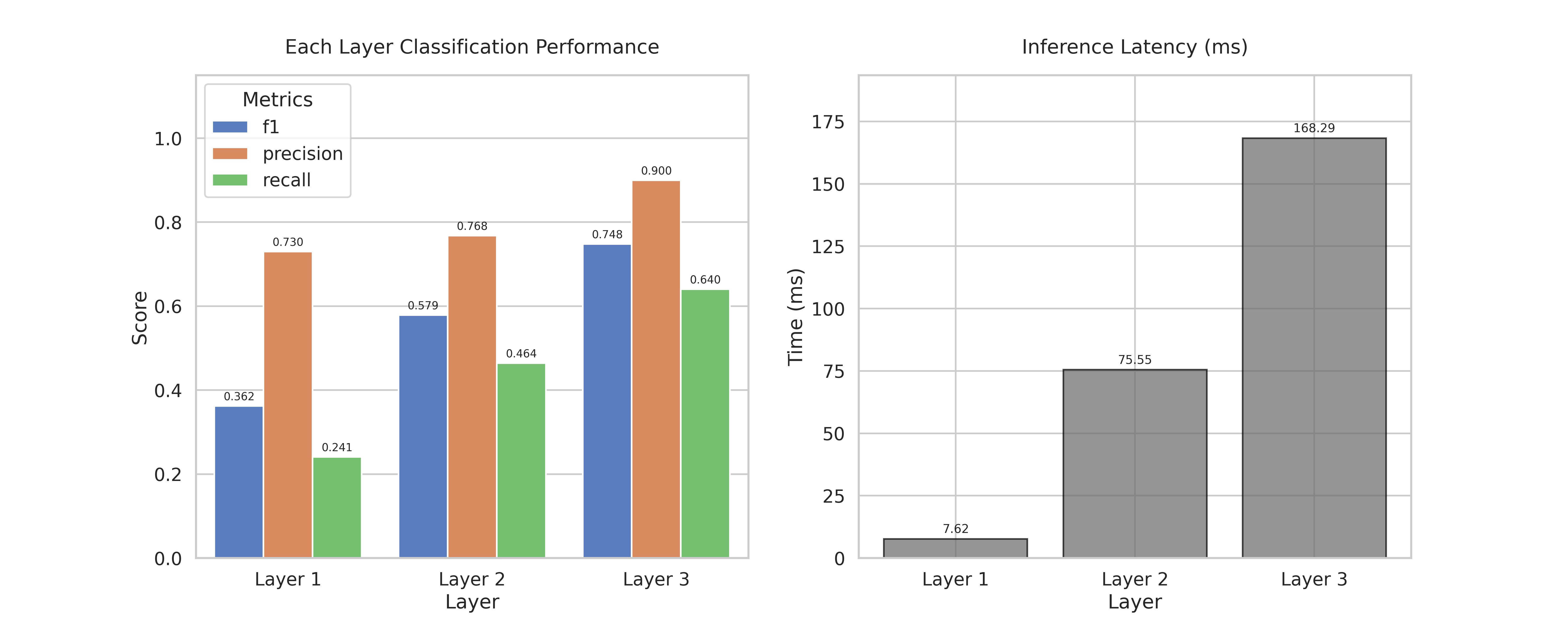}
    \caption{Standalone classification performance and inference latency of each defense layer.}
    \label{fig:each_layer}
\end{figure*}

This section provides additional visual evidence for the detection workflow and layer-level behavior. Figure~\ref{fig:problem_alt} gives an alternative view of the threat model, where the detector decides whether the untrusted context should be passed unchanged or sanitized. Figure~\ref{fig:example_run} shows a concrete C-auditing example in which a misleading identifier steers the model toward an unsafe answer, while neutralization restores the correct decision. Figure~\ref{fig:each_layer} complements the ablation in the main text by showing the standalone accuracy--latency trade-off of each layer.

\section{Transfer and Tokenizer Robustness}

\begin{figure*}[h]
    \centering
    \includegraphics[width=.9\linewidth]{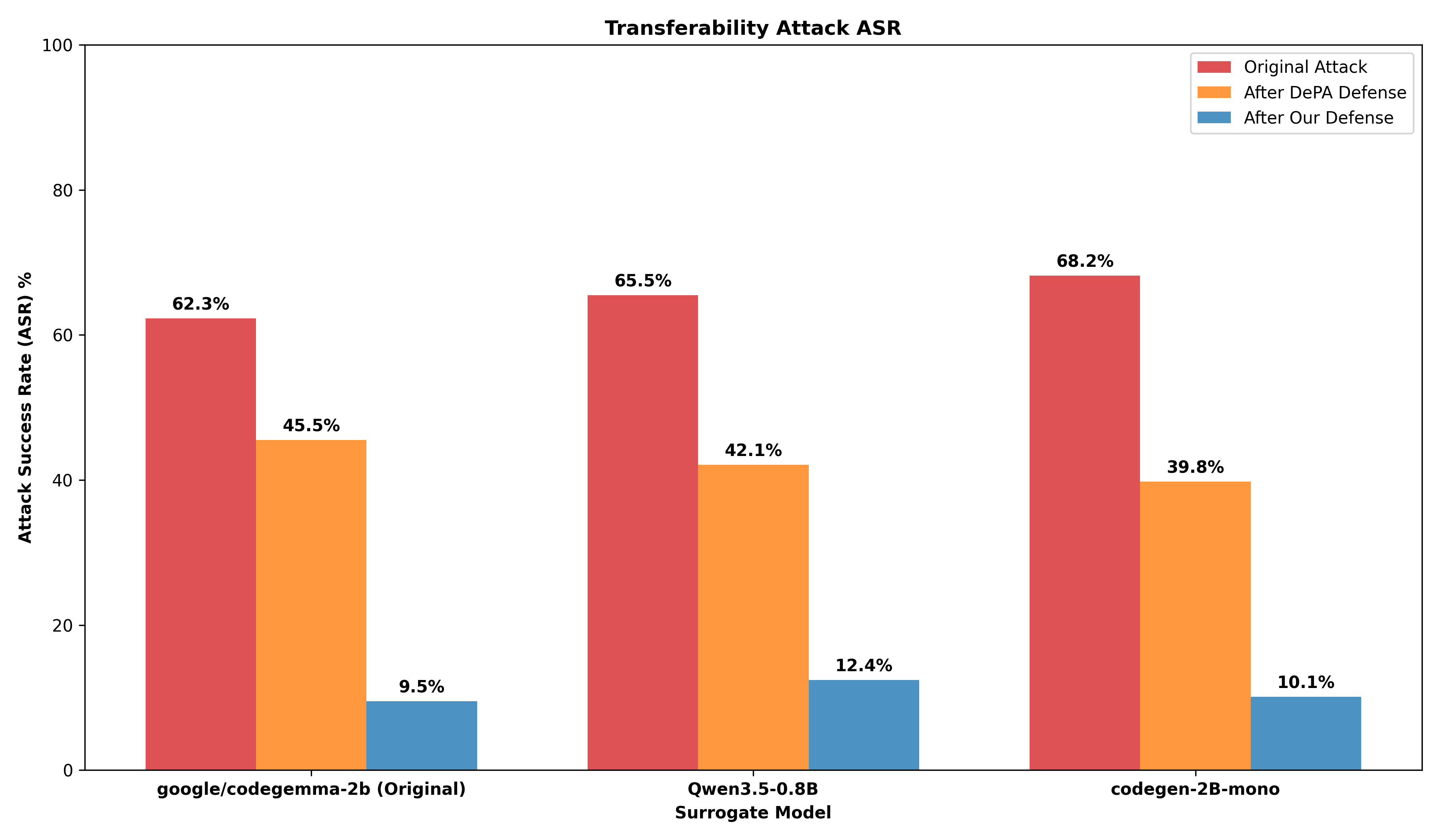}
    \caption{Transferability effect on sample-level ASR before and after defense.}
    \label{fig:transfer}
\end{figure*}

\begin{figure*}[h]
    \centering
    \includegraphics[width=.9\linewidth]{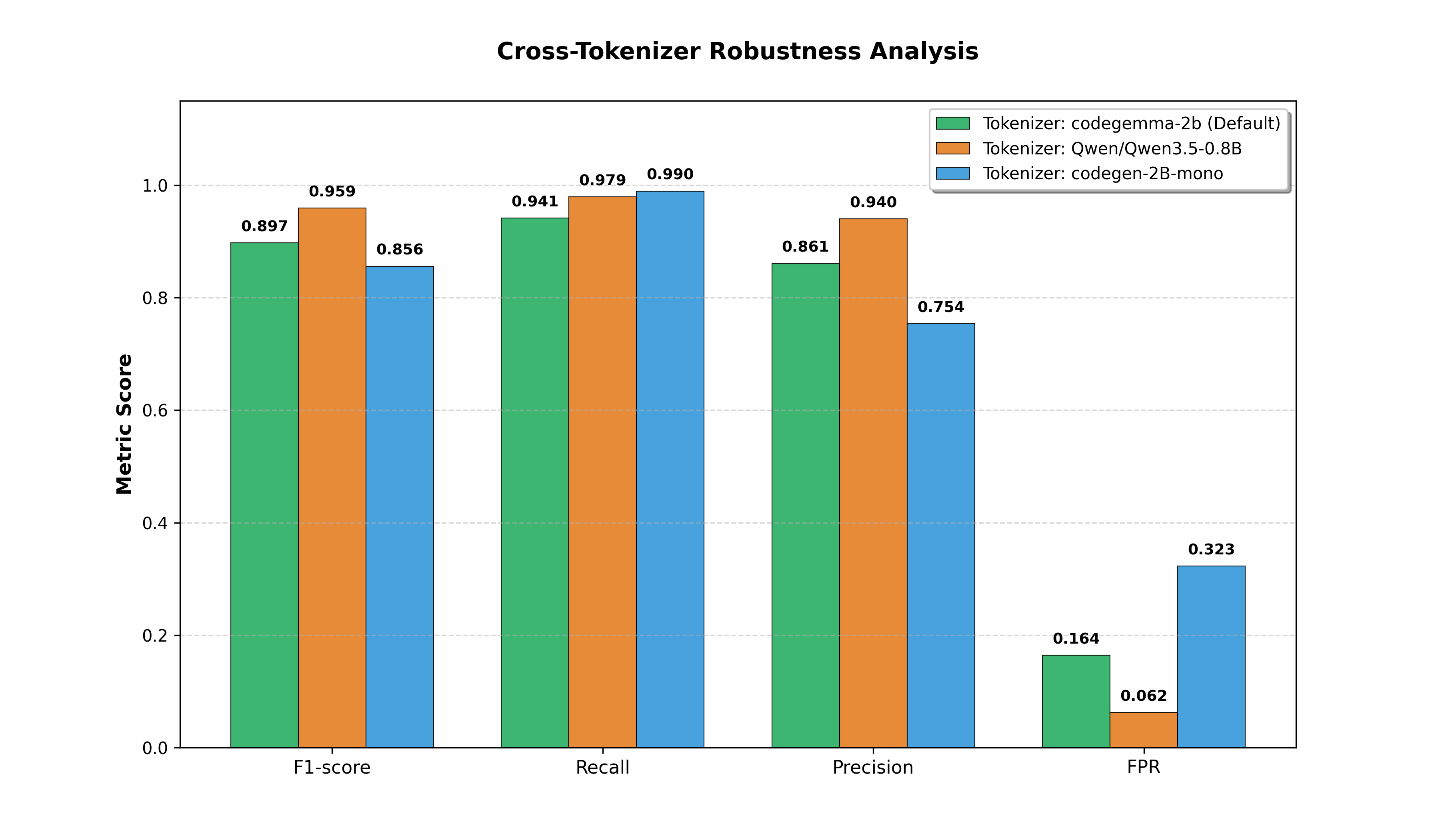}
    \caption{Cross-tokenizer robustness when attack payloads and defense detection use different tokenizer spaces. FPR denotes false positive rate; tokenizer space denotes the surrogate tokenizer used to segment payloads.}
    \label{fig:tokenizer}
\end{figure*}

We further test whether the detection signal transfers across victim models and tokenizers. Figure~\ref{fig:transfer} evaluates transferability by comparing the original attack success rate with ASR after DePA and after \method. The results show that \method reduces transferred attacks more consistently across victim models. Figure~\ref{fig:tokenizer} evaluates cross-tokenizer robustness. Since \method scores CST-aligned nodes rather than raw file-level text, its detection remains stable when the attack payload and defense-side surrogate use different tokenizer spaces.

Table~\ref{tab:depa_targeted} reports a targeted setting where attacks are optimized against DePA. The smaller F1 drop for \method suggests that CST-node localization and behavioral perturbation provide complementary signals beyond line-level perplexity.

\begin{table}[h]
\centering
\small
\begin{tabular}{lccc}
\toprule
Defense & Original F1 & Targeted F1 & Impact \\
\midrule
\method & 0.8061 & 0.7202 & -0.0859 \\
DePA & 0.5041 & 0.2300 & -0.2741 \\
\bottomrule
\end{tabular}
\caption{Detection performance under attacks optimized against the DePA baseline defense. Impact is computed as Targeted F1 minus Original F1.}
\label{tab:depa_targeted}
\end{table}

\section{Leave-One-Attack-Out Evaluation}

Table~\ref{tab:leave_one_out} evaluates whether \method depends on attack-family-specific signals. For each row, we remove one attack family during threshold selection and report performance on the held-out family. The average drop is modest, indicating that the pipeline captures general code-context injection patterns. The larger drops on XOXO and ITGen suggest that natural-looking semantic transformations remain the most challenging cases, which motivates the Layer-3 perturbation analysis.

\begin{table}[h]
\centering
\small
\begin{tabular}{lccc}
\toprule
Attack & Original & Leave-one-out & Impact \\
\midrule
XOXO & 0.80 & 0.673 & -0.127 \\
ITGen & 0.77 & 0.630 & -0.140 \\
Flashboom & 0.79 & 0.865 & 0.075 \\
CoTDeceptor & 0.66 & 0.660 & 0.000 \\
ShadowCode & 0.85 & 0.850 & 0.000 \\
INSEC & 0.90 & 0.900 & 0.000 \\
\midrule
Average & 0.80 & 0.763 & -0.037 \\
\bottomrule
\end{tabular}
\caption{Leave-one-attack-out evaluation. Impact is computed as Leave-one-out performance minus Original performance.}
\label{tab:leave_one_out}
\end{table}

\begin{table}[t]
\centering
\small
\begin{tabular}{lccc}
\toprule
Attack & Original & Adaptive & Impact \\
\midrule
Decoy Injection & 0.82 & 0.74 & -0.08 \\
Copy Trigger & 0.82 & 0.62 & -0.20 \\
Contextual Attack & 0.82 & 0.66 & -0.16 \\
\bottomrule
\end{tabular}
\caption{Node-level AUROC under adaptive attacks.}
\label{tab:adaptive}
\end{table}

\section{Adaptive Attack}
\label{app:adaptive_attack}
We further evaluate \method under an adaptive threat model.
In this setting, the attacker is assumed to know the defense
architecture and detection algorithms. In particular, the attacker
is aware that \method uses CST-guided candidate extraction,
surprise-gated selection in Layer~3, and node-level influence
measurement to detect semantic triggers. We consider three adaptive
strategies, each targeting a different component of the defense.

\paragraph{Decoy Injection.}
This attack targets the candidate-selection mechanism in Layer~3.
Because node perturbation analysis is more expensive than the earlier
layers, \method only applies it to selected high-risk candidates.
The attacker exploits this design by injecting high-surprise but
non-malicious nodes into the code context. These nodes are not intended
to directly cause malicious model behavior. Instead, they are designed
to occupy Layer-3 candidate slots, reducing the chance that the true
malicious trigger is evaluated by node perturbation analysis. In our
evaluation, the attacker injects 20 high-surprise non-malicious nodes
to stress this selection mechanism.

\paragraph{Copy Trigger.}
This attack targets the single-node influence assumption of Layer~3.
Instead of inserting one malicious trigger, the attacker places multiple
identical triggers at different locations in the code context. If
\method neutralizes only one occurrence, the remaining copies may still
preserve the attacker-desired behavior. As a result, the logits change
caused by neutralizing any single trigger can become smaller, making
each individual trigger appear less influential. This strategy tests
whether node perturbation analysis remains effective when the malicious
effect is distributed across multiple syntactic locations.

\paragraph{Contextual Attack.}
This attack targets the surprise-gated selector. The attacker uses a
genetic algorithm to optimize the surrounding context of the malicious
payload, with the goal of reducing the trigger's surprise score. If the
trigger appears less surprising under the defense-side scoring model, it
becomes harder to select for perturbation analysis. This evaluates
whether \method can still detect semantic triggers when the attacker
explicitly optimizes the local context to avoid the selection stage.

\paragraph{Results.}
Table~\ref{tab:adaptive} reports the F1 score under the three adaptive attacks. The original defense achieves an F1 score of 0.82. Under Decoy Injection, the F1 score drops to 0.74, showing that high-surprise benign nodes can partially distract the candidate selector. Under Copy Trigger, the F1 score drops to 0.62, which is the largest degradation among the three attacks. This confirms that duplicated triggers can dilute the influence of any single neutralized node. Under Contextual Attack, the F1 score decreases to 0.66, indicating that reducing the trigger's surprise score weakens the perturbation-selection stage. Overall, \method degrades under adaptive pressure but still maintains an F1 score of at least 0.62.

The adaptive evaluation highlights two limitations of influence-based semantic-trigger detection. First, high-surprise benign nodes can consume part of the perturbation budget. Second, duplicated triggers can reduce the apparent influence of any single occurrence. Nevertheless, \method does not collapse under these stronger attacks, suggesting that syntax-aware localization, anomaly scoring, and node-level perturbation provide complementary robustness.

\end{document}